\documentclass[12pt,preprint]{aastex}

\shorttitle{Rossby}
\shortauthors{Zaqarashvili et al.}

\begin{document}

\title{Quasi-biennial oscillations in the solar tachocline caused by magnetic Rossby wave instabilities}

\author{Teimuraz V. Zaqarashvili\altaffilmark{1,4},
Marc Carbonell\altaffilmark{2}, Ram\'{o}n Oliver\altaffilmark{3}, and Jos\'{e} Luis
Ballester\altaffilmark{3}}

\altaffiltext{1}{Space Research Institute, Austrian Academy of Sciences,
Schmiedlstrasse 6, 8042 Graz, Austria. Email: teimuraz.zaqarashvili@oeaw.ac.at}
\altaffiltext{2}{Departament de Matem\`{a}tiques i Inform\`{a}tica. Universitat de les Illes Balears, \\ E-07122 Palma de Mallorca,
Spain. Email: marc.carbonell@uib.es} \altaffiltext{3}{Departament de
F\'{\i}sica, Universitat de les Illes Balears, E-07122 Palma de
Mallorca, Spain. Email: ramon.oliver@uib.es,
joseluis.ballester@uib.es} \altaffiltext{4}{Abastumani Astrophysical Observatory at
Ilia State University, Chavchavadze Ave 32, 0179
Tbilisi, Georgia.}

\begin{abstract}
Quasi-biennial oscillations (QBO) are frequently observed in the
solar activity indices. However, no clear physical mechanism for the
observed variations has been suggested so far. Here we study the
stability of magnetic Rossby waves in the solar tachocline using the
shallow water magnetohydrodynamic approximation. Our analysis shows
that the combination of typical differential rotation and a toroidal
magnetic field with a strength $\geq 10^5$ G triggers the
instability of the $m=1$ magnetic Rossby wave harmonic with a period
of $\sim 2$ years. This harmonic is antisymmetric with respect to
the equator and its period (and growth rate) depends on the
differential rotation parameters and the magnetic field strength.
The oscillations may cause a periodic magnetic flux emergence at the
solar surface and consequently may lead to the observed QBO in the
solar activity features. The period of QBO may change throughout the
cycle, and from cycle to cycle, due to variations of the mean
magnetic field and differential rotation in the tachocline.


\end{abstract}


\section{Introduction}\label{intro}

Apart from the well known 11-year cycle, solar activity shows quasi
periodic variations on shorter time scales. Two different time
scales have been frequently observed in many solar activity
indicators: several months and a few years. The oscillations with
period of several months (mostly with 150-160 days) are known as
Rieger-type periodicities \citep{rieger84, lean89, carball90,
oliver98, krivo02, kane05, knaack05}. The oscillations with period
$\sim$ 2 years are known as Quasi-Biennial Oscillations (QBO) and
they modulate almost all indices of solar activity \citep{sakurai81,
gigolashvili95, knaack05, kane05, danilovic05, forgacs07,
badalyan08, javaraiah09, laurenza09, vecchio09, vecchio10,
sykora10}.

The source(s) of these periodicities is still unclear. Several
mechanisms have been suggested to drive the Rieger-type
periodicities: interaction between $l=2$ and $l=3$ g-modes
\citep{wolff83}, the timescale for storage and/or escape of magnetic
fields in the solar convection zone \citep{Ichi85}, \lq \lq clock"
modeled by an oblique rotator \citep{Bai91} and equatorially trapped
Rossby-type waves in the photosphere \citep{lou00}. Recently,
\citet{zaqa10} (hereinafter Paper I) suggested that the Rieger-type
periodicities can be caused by unstable $m=1$, two-dimensional
($\theta- \phi$ surface in spherical coordinates) magnetic Rossby
waves in the solar tachocline. They show that a combination of the
typical differential rotation parameters and the magnetic field
strength $\leq 10^4$ G in the tachocline favor the strong growth of
one particular harmonic with period of 150-160 days. The periodic
modulation of the tachocline magnetic field due to the unstable
harmonic triggers the periodic emergence of magnetic flux towards
the surface, which leads to the observed periodicities in the solar
activity. On the other hand, there is no clear mechanism for QBO
reported in literature so far. \citet{pataraya95} supposed that the
quasi 2-year impulse of shear waves can cause the 2-year periodicity
of the differential rotation in the photosphere. However, this
mechanism may work only near solar minima and cannot explain the
long standing modulation of solar activity. Therefore, the question
of the source for QBO is widely open.

In this letter, we show that the instability of magnetic Rossby
waves in the tachocline could be the reason for QBO in solar
activity. We consider a nonzero thickness of the tachocline and
hence we use the shallow water magnetohydrodynamic (SWMHD) equations
\citep{gil00}. We show that a stronger magnetic field ($\geq 10^5$
G) favors the growth of magnetic Rossby wave harmonic with period
$\sim$ 2 years.

\section{Shallow water MHD equations and unstable magnetic Rossby wave harmonics}

In the following we use a spherical coordinate system $(r, \theta,
\phi)$  rotating with the solar equator, where $r$ is the radial
coordinate, $\theta$  is the co-latitude and $\phi$ is the
longitude.

The solar differential rotation law in general is
\begin{equation}\label{omega}
\Omega=\Omega_0 (1 - s_2 \cos^2 \theta -s_4 \cos^4 \theta)=\Omega_0 + \Omega_1(\theta),
\end{equation}
where $\Omega_0$ is the equatorial angular velocity, and $s_2, s_4$
are constant parameters determined by observations.

In the solar tachocline the magnetic field is predominantly
toroidal, $\vec B = \Xi \hat{e}_{\phi}$, and we take $\Xi =
B_{\phi}(\theta) \sin \theta$, where $B_{\phi}$ is in general a
function of co-latitude. Then, the linear SWMHD equations
\citep{gil00} can be rewritten in the rotating frame (with
$\Omega_0$) as follows:
\begin{equation} \label{MHD1}
{{\partial u_{\theta}}\over {\partial t}} +
\Omega_1{{\partial u_{\theta}}\over {\partial \phi}} - 2\Omega\cos \theta u_{\phi} = -{{g}\over
{R_0}}{{\partial h}\over {\partial \theta}}+ {{B_{\phi}}\over
{{4\pi\rho R_0}}}{{\partial b_{\theta}}\over {\partial \phi}}-2
{{B_{\phi} \cos \theta}\over {{4\pi\rho R_0 }}}b_{\phi},
\end{equation}
\begin{equation}\label{MHD2}
{{\partial u_{\phi}}\over {\partial t}} +
\Omega_1{{\partial u_{\phi}}\over {\partial \phi}}  +\frac{u_{\theta}}{\sin{\theta}}{{\partial [\Omega\sin^2{\theta}]}\over
{\partial \theta}}=-{{g}\over {R_0
\sin \theta}}{{\partial h}\over {\partial \phi}}+{{B_{\phi}}\over
{{4\pi\rho R_0}}}{{\partial b_{\phi}}\over {\partial \phi}}+
{{b_{\theta}}\over {{4\pi\rho R_0 \sin \theta}}}{{\partial [B_{\phi}\sin^2 \theta]}\over
{\partial \theta}},
\end{equation}
\begin{equation}\label{MHD3}
{{\partial b_{\theta}}\over {\partial t}} +
\Omega_1{{\partial b_{\theta}}\over {\partial \phi}} = {{B_{\phi}}\over {{R_0}}}{{\partial u_{\theta}}\over
{\partial \phi}},
\end{equation}
\begin{equation}\label{MHD4}
{{\partial }\over {\partial \theta}}\left (\sin \theta b_{\theta}
\right ) + {{\partial b_{\phi}}\over {\partial
\phi}}+\frac{B_{\phi}\sin \theta}{H_0}{{\partial h}\over {\partial
\phi}}=0,
\end{equation}
\begin{equation}\label{MHD5}
{{\partial h}\over {\partial t}} +
\Omega_1{{\partial h}\over {\partial \phi}}
+\frac{H_0}{R_0 \sin{\theta}}{{\partial }\over {\partial
\theta}}\left (\sin \theta u_{\theta} \right ) + \frac{H_0}{R_0
\sin{\theta}}{{\partial u_{\phi}}\over {\partial \phi}}=0,
\end{equation}
where $u_{\theta}$, $u_{\phi}$, $b_{\theta}$ and $b_{\phi}$ are the
velocity and magnetic field perturbations, $H_0$ is the tachocline
thickness and $h$ is its perturbation, $\rho$ is the density, $g$ is the reduced gravity and
$R_0$ is the distance from the solar center to the tachocline. Eqs. (\ref{MHD4})-(\ref{MHD5}) are the solenoidal conditions of shallow water magnetic field and velocity respectively \citep{gil00}.

Fourier analysis with $\exp[im(\phi - c t)]$ and the transformation
of variables $\mu=\cos \theta$ in Eqs. (\ref{MHD1})-(\ref{MHD5}) lead to the equations
\begin{equation}\label{working1}
[(c-\Omega_1)^2 -\Omega^2_A](1-\mu^2)\frac{\partial H}{\partial \mu} -
2\mu[\Omega(c-\Omega_1) +\Omega^2_A] H=-im
\Omega^2_gh+im(1-\mu^2)[(c-\Omega_1)^2-\Omega^2_A]h,
\end{equation}
$$
2\mu[\Omega(c-\Omega_1) +\Omega^2_A](1-\mu^2)\frac{\partial H}{\partial \mu}
- \left [m^2(c-\Omega_1)^2 -m^2\Omega^2_A+\mu(1-\mu^2)\frac{\partial
\Omega^2}{\partial \mu} -\mu (1-\mu^2)\frac{\partial \Omega^2_A}{\partial
\mu} \right ]H=
$$
\begin{equation}\label{working2}
=im\Omega^2_g(1-\mu^2)\frac{\partial h}{\partial \mu}
+2im\mu(1-\mu^2)[\Omega(c-\Omega_1)+\Omega^2_A]h,
\end{equation}
where
$$
\Omega_A=\frac{B_{\phi}}{\sqrt{4 \pi \rho} R_0},\,\,
\Omega_g=\frac{\sqrt{gH_0}}{R_0},
$$
are the Alfv\'en and surface gravity frequency respectively, $h$ is
normalized by $H_0$ and
\begin{equation}\label{H}
H(\mu)=\frac{b_{\theta}(\mu)\sqrt{1-\mu^2}}{B_{\phi}}=\frac{u_{\theta}(\mu)\sqrt{1-\mu^2}}{R_0(c-\Omega_1)}.
\end{equation}

In the remaining we use a magnetic field
\begin{equation}\label{magnetic}
B_{\phi}=B_0 \mu,
\end{equation}
which changes sign at the equator \citep{gilman97}.

We expand $H$ and $h$ in infinite series of associated
Legendre polynomials \citep{longuet68}
\begin{equation}\label{legandre}
H=\sum^{\infty}_{n=m}a_nP^m_n(\mu),\,\,\,h=\sum^{\infty}_{n=m}b_nP^m_n(\mu),
\end{equation}
which satisfy the boundary conditions $H=h=0$ at $\mu=\pm 1$ (i.e.
at the solar poles). Then, we substitute Eq. (\ref{legandre}) into
Eqs. (\ref{working1})-(\ref{working2}) and, using a recurrence
relation of Legendre polynomials, we obtain algebraic equations as
infinite series. The dispersion relation for the infinite number of
harmonics can be obtained when the infinite determinant of the
system is set to zero. In order to solve the determinant, we
truncate the series at $n=70$, and the resulting polynomial in
$\omega$ is solved numerically. This gives the frequencies of the
first 70 harmonics. The harmonics with real frequency are stable,
while those with complex frequency are unstable (see the general
technique of Legendre polynomial expansion in
\citet{longuet68,watson81,gilman97,zaqa10} and references therein).

The typical values of equatorial angular velocity, radius and
density in the tachocline are $\Omega_{0}= 2.7 \cdot 10^{-6}$
s$^{-1}$, $R_{0} = 5 \cdot 10^{10}$ \ cm and $\rho = 0.2$ \ g
$\cdot$ cm$^{-3}$ respectively. We use a tachocline thickness
$H_0=0.02 R_0=10^{9}$ \ cm. The ratio between angular and surface
gravity frequencies $\epsilon= \Omega^2_{0}/\Omega^2_g= \Omega^2_{0}
R^2_{0}/(gH_0)$ is an important parameter in the shallow water
theory. $\epsilon \ll 1$ means  strongly stable stratification (main
part of tachocline), while $\epsilon \gg 1$ considers weakly stable
stratification (upper overshoot region). Here we consider the mean
part of the tachocline and thus we take the limit $\epsilon \ll 1$.

The observed differential
rotation parameters near the solar surface satisfy $s_{2}+ s_{4}
\approx 0.28$, which may tend to $s_{2}+ s_{4} \approx 0.26$ near
the upper part of the tachocline \citep{schou98}. The solar
radiative interior rotates uniformly, therefore the latitudinal
differential rotation parameters drop to zero from the upper part of
tachocline to its base. The radial dependence of latitudinal
differential rotation through the tachocline is not clear, and
$s_2, s_4$ may also vary throughout the solar cycle \citep{howe2000}.
Therefore, $s_2+s_4$ may take any value between 0.26 and 0.

\begin{figure}
    \epsscale{1.2}
\plotone{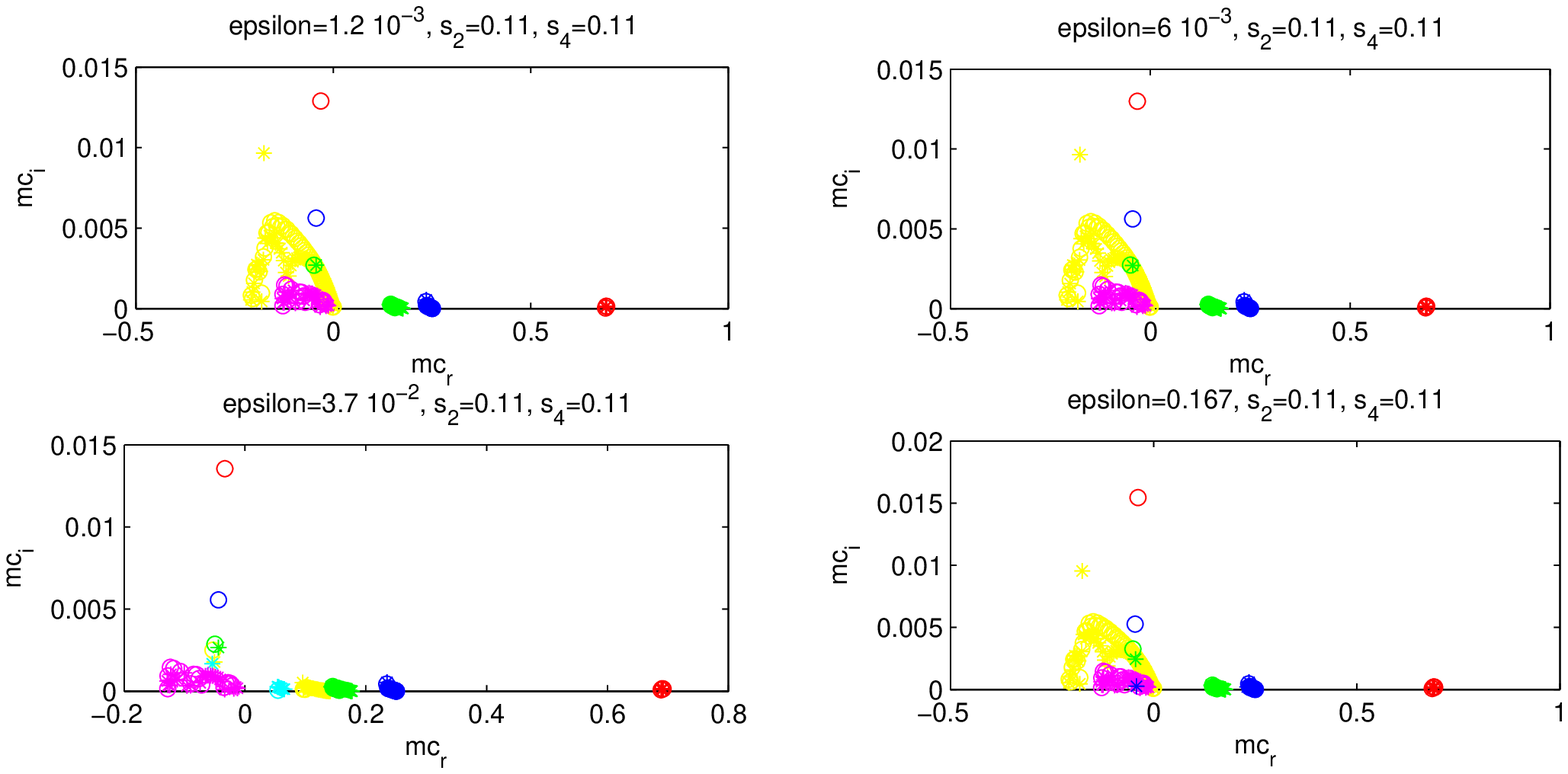}
 \caption{Real, $mc_r$, and imaginary, $mc_i$,
frequencies of all $m=1$ unstable harmonics for different
combinations of $\epsilon$ (i.e. reduced gravity) and magnetic field
strength. The differential rotation parameters are fixed to
$s_2=s_4=0.11$ for all four panels. The reduced gravity varies from
$1.5 \cdot 10^4$ cm s$^{-2}$  ($\epsilon=1.2 \cdot 10^{-3}$, upper
left panel) to $1.1 \cdot 10^2$ cm s$^{-2}$ ($\epsilon=0.167$, lower
right panel). Yellow, magenta, blue, green, dark blue and red colors
correspond to magnetic field strengths of $2 \cdot 10^{3}$ \ G, $2
\cdot 10^{4}$ \ G, $6 \cdot 10^{4}$ \ G, $8 \cdot 10^{4}$ \ G,
$10^{5}$ \ G, and $2 \cdot 10^{5}$ \ G, respectively.  Asterisks
(circles) denote the symmetric (antisymmetric) harmonics with
respect to the equator. } \label{fig1}
\end{figure}

Figure~\ref{fig1} shows the real, $mc_r$, and imaginary, $mc_i$,
frequencies of all $m=1$ unstable harmonics for different
combinations of $\epsilon$ (i.e. reduced gravity) and magnetic field
strength.  The differential rotation parameters are fixed to
$s_2=s_4=0.11$.
The plot shows that each combination of the magnetic field strength,
differential rotation parameters and reduced gravity leads to the
occurrence of a particular harmonic whose growth rate is much
stronger than that of other harmonics. This is similar to what
happens in the two-dimensional case \citep{zaqa10}. An increase of
magnetic field strength leads to the reduction of the frequency of
the most unstable harmonic. The unstable harmonics are mostly
symmetric (defined by asterisks) with respect to the equator for a
magnetic field strength $< 10^{4}$ \ G, while they become mostly
antisymmetric (defined by circles) for a strength $> 10^{5}$ \ G. A
magnetic field strength between $10^{4}- 10^{5}$ \ G yields unstable
harmonics for both symmetries. This can be explained in terms of
magnetic and differential rotation energies. Equipartition between
the magnetic energy and the kinetic energy of differential rotation
occurs at $\sim 5 \cdot 10^{4}$ \ G for $s_2=s_4=0.11$. When the
magnetic field strength is smaller, then the differential rotation
is the main energy source for instability and this obviously yields
the symmetric harmonics as the differential rotation is symmetric
around the equator. However, when the magnetic field is stronger,
then the magnetic energy is the main source for the instability and
the unstable harmonics are antisymmetric as the magnetic field is
antisymmetric with respect to the equator.

The importance of the equipartition value of the magnetic field
strength is clearly seen on Figure~\ref{fig2}. This figure displays
the periods and growth rates (defined as $mc_i/\Omega_0$) vs
magnetic field strength. The growth rates are higher for weaker
($<10^4$ G) and stronger ($>10^5$ G) magnetic fields. However, the
growth rates are much lower when the magnetic field strength is
inside the interval $10^4-10^5$ G. The weaker field ($<10^4$ G)
favors Rieger-type periodicities (150-160 days), while the stronger
field ($>10^5$ G) supports QBO.  Increasing the magnetic field
suppresses symmetric harmonics as it has been shown in Paper I.

Figure~\ref{fig3} displays the period of the most unstable symmetric
and antisymmetric harmonic vs the value of reduced gravity (i.e. on
$\epsilon$) for a magnetic field strength of 8$\cdot 10^4$ \ G and
the differential rotation parameters $s_2=s_4=0.11$. It is seen that
the oscillation period does not depend significantly on the reduced
gravity.

Figure~\ref{fig4} displays the dependence of the periods of the most
unstable symmetric and antisymmetric harmonics on the differential
rotation parameters for a magnetic field strength of 8$\cdot 10^4$ \
G and for $\epsilon=0.12$ (corresponding to a reduced gravity of
$1.5 \cdot 10^2$ cm s$^{-2}$). The upper panel (circles) displays
the antisymmetric harmonics and the lower panel (asterisks) displays
the symmetric ones. The periods of unstable harmonics vs $s_4$ are
plotted for different values of $s_2$. The values of $s_2$ vary from
0.13 (blue color) to 0.09 (red color). We can observe that the
period of this harmonic depends  on the differential rotation
parameters significantly and takes values between 400-700 days. The
period becomes shorter for stronger differential rotation.

\section{Discussion}

\begin{figure}
\epsscale{0.8} \plotone{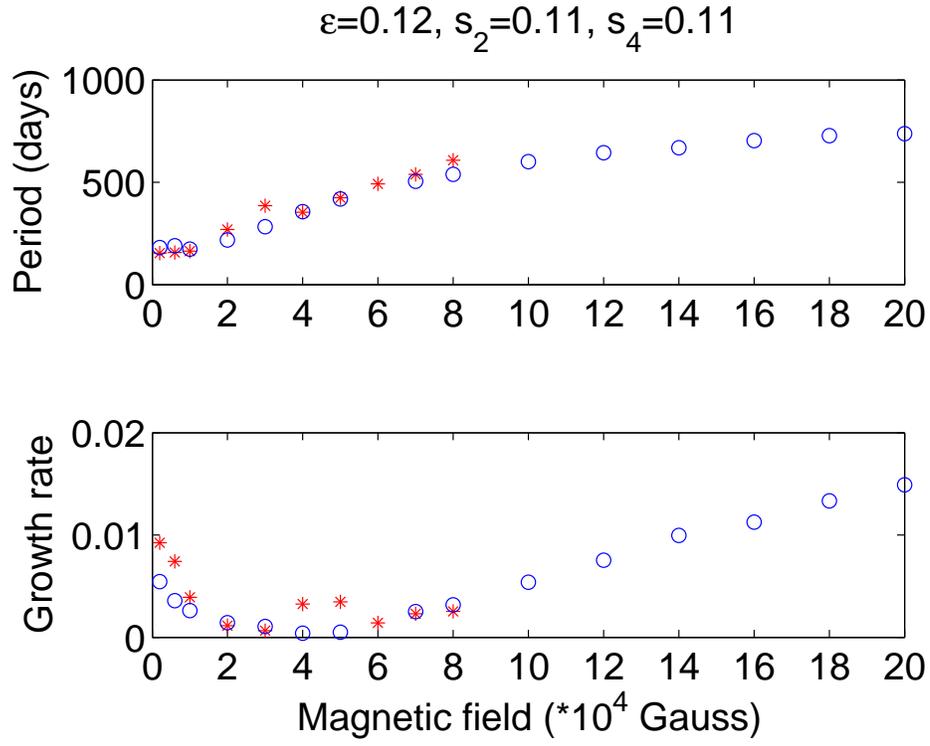} \caption{Period (upper panel) and
growth rate $mc_i/\Omega_0$ (lower panel) of the most unstable harmonics vs magnetic field
strength. Asterisks (circles) define symmetric (antisymmetric) harmonics. Here $\epsilon$=0.12 and the differential rotation parameters are $s_2=s_4=0.11$.} \label{fig2}
\end{figure}

\begin{figure}
\epsscale{1.0} \plotone{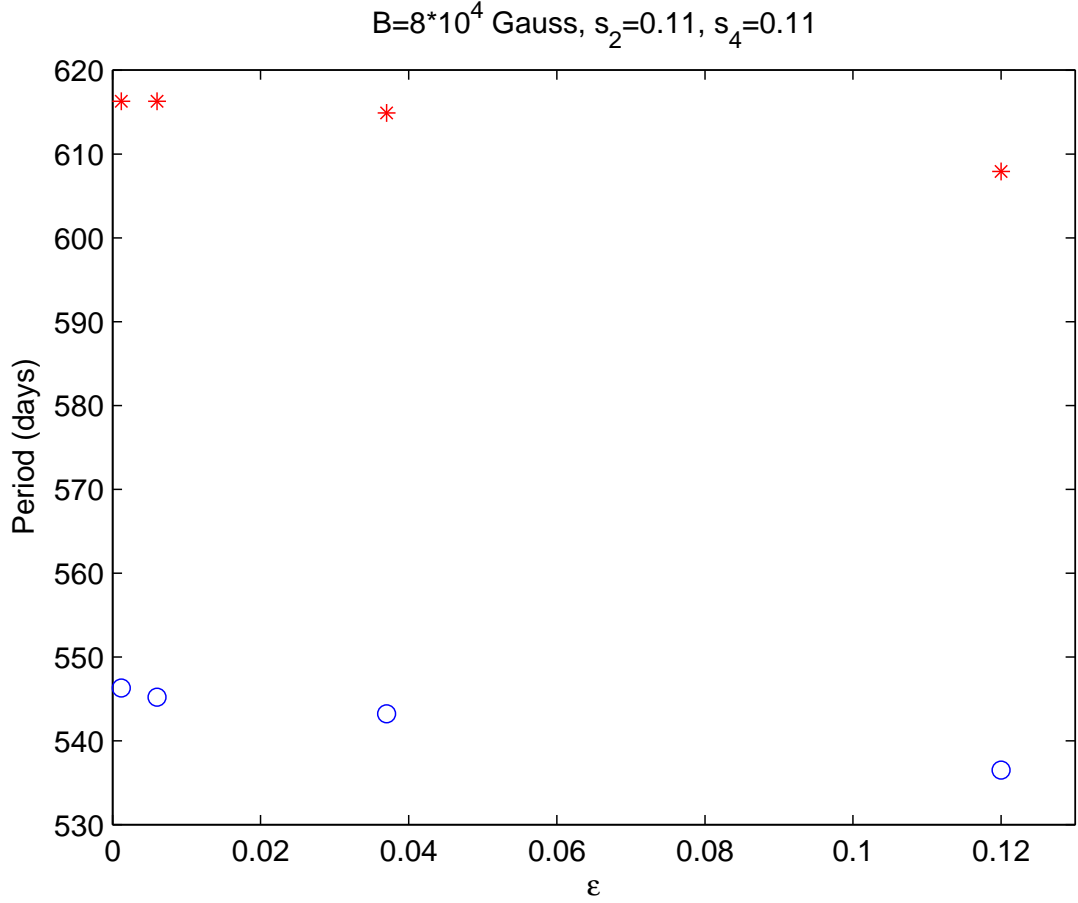} \caption{Periods of the most
unstable symmetric (asterisks) and antisymmetric (circles) harmonics
vs $\epsilon$ for a magnetic field strength of 8$\cdot 10^4$ \ G and
the differential rotation parameters $s_2=s_4=0.11$.}\label{fig3}
\end{figure}

\begin{figure}
\epsscale{1.0} \plotone{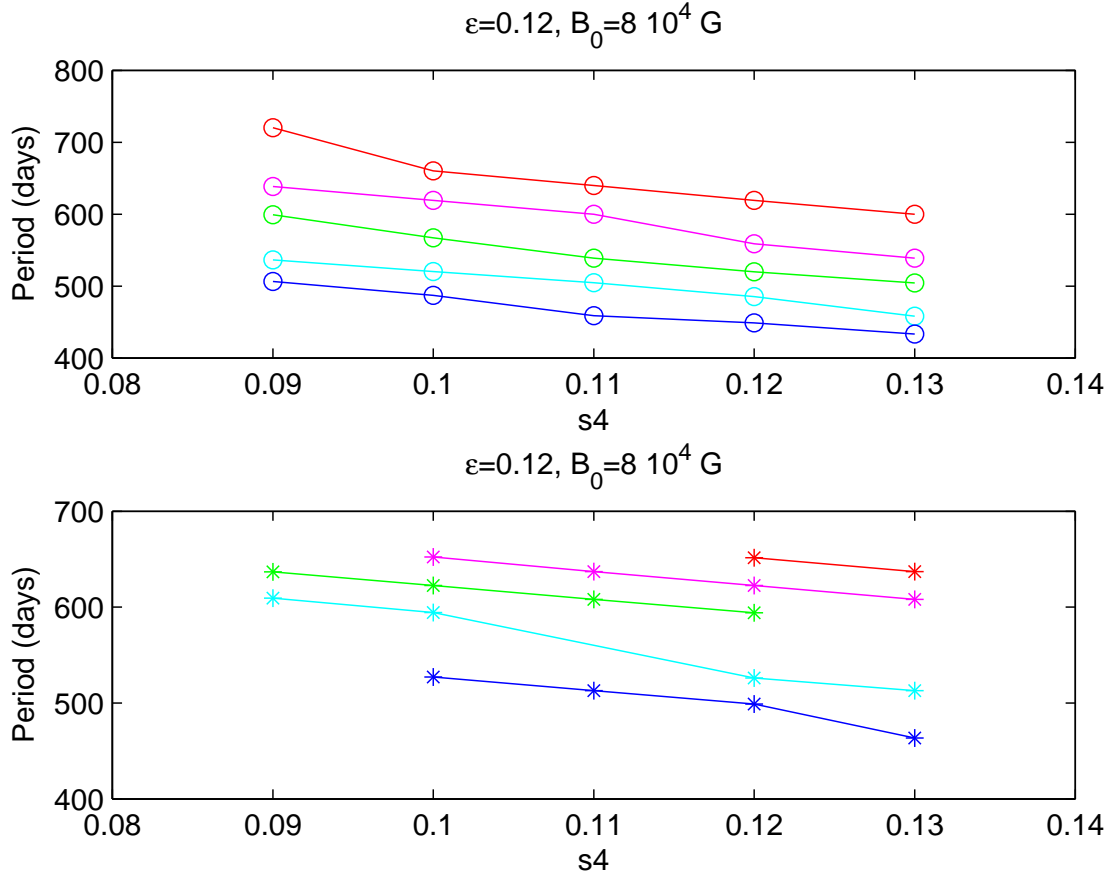} \caption{Periods of the most
unstable symmetric (asterisks) and antisymmetric (circles) harmonics
vs $s_4$ for different values of $s_2$. Dark blue, blue, green,
magenta and red colors correspond to $s_2$=0.13, 0.12, 0.11, 0.10
and 0.09 respectively. The magnetic field strength equals 8$\cdot
10^4$ \ G and $\epsilon=0.12$.} \label{fig4}
\end{figure}

Our results show that the differential rotation and the magnetic
field with a strength of $>10^5$ G may lead to large-scale
oscillations of tachocline with periods of $\sim$ 2 years. The
oscillation is due to the $m=1$ unstable mode of magnetic Rossby
waves. The magnetic Rossby waves are magnetohydrodynamic
counterparts to usual hydrodynamic Rossby waves
\citep{zaqa07,zaqa09}. The period and growth rate of the unstable
harmonics depend on the magnetic field strength and the differential
rotation parameters (Figures ~\ref{fig1}, ~\ref{fig2} and
~\ref{fig4}). The unstable harmonics with periods of $\sim 2$ years
are antisymmetric with respect to the solar equator.

The unstable magnetic Rossby waves in the tachocline can be the
reason for QBO observed in almost all indices of the solar activity.
Recent papers suggest that QBO are not persistent but may vary from
cycle to cycle \citep{vecchio09} and throughout a cycle
\citep{sykora10}. Our analyses also suggest this behaviour as the
period of unstable harmonics depends on magnetic field strength and
differential rotation parameters, which may vary in time depending
on phase and strength of a particular cycle.

The antisymmetric behaviour of unstable harmonics with respect to
the equator may explain the recent observational results of
\citet{badalyan08}, which show that QBO are well recognizable in the
N-S asymmetry of solar activity indices.


Our analysis suggests the reduction of growth rates of unstable
harmonics when the magnetic field strength is inside the interval
$10^4-10^5$ G (see Figure ~\ref{fig2}). It is clearly seen that the
relatively weak magnetic field $<10^4$ G leads to the occurrence of
Rieger-type periodicities (see the same results in the paper I),
while the field of $>10^5$ G favors QBO. The upper overshoot region
of the tachocline probably contains relatively weaker magnetic field
comparing to the lower stable layers. Therefore, we may speculate
that the Rieger-type periodicities are formed in the overshoot layer
(this was also suggested in the paper I), while QBO are formed in
lower layers with strongly stable stratification. Therefore, the
both periodicities may occur simultaneously.

The magnetic field of $10^5$ G is unstable due to the buoyancy
instability, which makes difficult to keep it in the tachocline. On
the other hand, the emergence of magnetic flux at observed latitudes
requires sufficiently strong magnetic field ($\sim 10^5$) below the
convection zone \citep{Fan04}. Therefore, the storage of strong
fields below the convection zone is still open question.

Significant simplifications in our approach is the linear stability
analysis, which only describes the initial phase of instability.
Intense numerical simulations are needed in the future to study the
developed stage of magnetic Rossby wave instabilities.

%

\section{Conclusions}

We have shown that the unstable magnetic Rossby waves in the solar
tachocline could be responsible for the observed intermediate
periodicities in solar activity. The periods and growth rates of
unstable harmonics depend on the differential rotation parameters
and the magnetic field strength. The unstable harmonics are either
symmetric or antisymmetric with respect to the equator. The
latitudinal differential rotation is mainly responsible for the
growth of symmetric harmonics, while, the antisymmetric toroidal
magnetic field favors the growth of antisymmetric harmonics. A
magnetic field with a strength $\leq 10^4$ G leads to oscillations
with shorter period (150-170 days), while a stronger magnetic field
$\geq 10^5$ G favors oscillations with longer periods (1-2.5 yrs).
Hence, $\sim$ 2-year oscillations can be formed in the main part of
the tachocline with stronger toroidal magnetic field and strongly
stable stratification. The oscillations may trigger the periodic
magnetic flux emergence at the solar surface and consequently QBO in
solar activity.




{\bf Acknowledgements} The authors acknowledge the financial support
provided by MICINN and FEDER funds under grant AYA2006-07637.
T. V. Z. acknowledges
financial support from the Austrian Fond zur F\"orderung der wissenschaftlichen Forschung (under project P21197-N16), the Georgian National Science Foundation (under grant GNSF/ST09/4-310) and the Universitat de les Illes
Balears.

\clearpage

\clearpage



\end{document}